# Nanoscale Mechanical Structures Fabricated from Silicon-on-Insulator Substrates

A.N. Cleland and M.L. Roukes
*M.S. 114-36, California Institute of Technology,
Pasadena CA 91125*

**Abstract**

We describe a method with which to fabricate sub-micron mechanical structures from silicon-on-insulator substrates. We believe this is the first reported method for such fabrication, and our technique allows for complex, multilayer electron beam lithography to define metallized layers and structural *Si* layers on these substrates. The insulating underlayer may be removed by a straightforward wet processing step, leaving suspended single crystal *Si* mechanical structures. We have fabricated and mechanically tested structures such as beam resonators, tuning-fork resonators, and torsional oscillators, all with smallest dimensions of 0.1-0.2 μm and fundamental resonance frequencies above $10^7$ Hz.

The recent advent of easily produced silicon-on-insulator substrates greatly simplifies the fabrication of suspended nanoscale mechanical structures. In this paper we describe a method we have developed to fabricate submicron, very high frequency mechanical resonators using SIMOX (separation by implantation of oxygen) substrates, such that the structures have minimum dimensions of 0.1-0.2 μm, and potential resonance frequencies up to 1 GHz; we have measured fundamental resonances up to 100 MHz.

Other authors have reported recipes for fabricating sub-micron suspended *Si* structures[1-3]. These recipes were however somewhat complicated and all suffered from an inability to precisely control the thickness of the suspended structures. In our process the thickness is defined by the substrate *Si* top layer thickness, which is very well controlled in the SIMOX process. The authors of Ref. 4 used silicon-on-insulator substrate produced by wafer bonding, where the top *Si* layer had a thickness of 6 μm; to achieve the very high frequencies and small dimensions described here, a reduction of this thickness by at least an order of magnitude is required.

We first describe the recipe used to fabricate the structures, and then describe the method used to measure the resonance properties of suspended submicron *Si* resonators, showing data for a few different structures.

The substrates we used were SIMOX <110> orientation n-type *Si* wafers[5], which included a 400 nm thick buried $SiO_2$ layer with a 200 nm thick top layer of single-crystal *Si*. Lithographic processing was performed in four steps, consisting of an optically defined alignment layer, an scanning electron microscope (SEM) lithographically defined wiring layer, a contact pad definition layer, and an SEM-defined metal masking layer for reactive plasma etching of the *Si*.

Following the *Si* etch, a wet etch was used to remove the exposed regions of buried oxide, which resulted in mechanical suspension of the *Si* structures.

We cleaned the substrates by soaking in detergent, acetone and methanol. The first alignment layer was defined by optical lithography followed by a 5 nm *Cr*/30 nm *Au* thermal evaporation, with liftoff in acetone. To define the detailed circuit wiring, we deposited a PMMA bilayer on the substrates, with a 400 nm thick underlayer of 496 KD PMMA and a 100 nm thick overlayer of 960 KD PMMA. We patterned the bilayer by SEM lithography with a 40 kV accelerating voltage and a dose of 400-500 $\mu C/cm^2$; the pattern was developed in 1 part 4-methyl-2-pentanone:3 parts isopropanol, with an isopropanol rinse. A 5 nm *Cr*/30 nm *Au* layer was deposited by thermal evaporation and liftoff carried out in heated acetone; the substrates then had the appearance shown in Fig. 1(a).

Large area contacts were made to the previously defined layer by optical lithography, followed by a 8 nm *Cr*/120 nm *Au* thermal evaporation. We then deposited 120 nm of *Ni* by RF sputter deposition, followed by liftoff in acetone; the *Ni* serves as a mask for reactive ion etching of the *Si*. The detailed mechanical structure was then defined by SEM lithography using the parameters described above, followed by sputter deposition of 100 nm of *Ni*. The substrates then had the appearance shown in Fig. 1(b).

The pattern in the *Ni* mask was transferred to the substrate by anisotropic reactive ion etch in a parallel plate reactive ion etcher, which has a 6" dia. cathode, a 4" anode-to-cathode spacing and operates at 13.56 MHz. The cathode plate is water cooled to a little below room temperature. The etch chamber is oil diffusion pumped, and was always pumped to better than $2 \times 10^{-5}$ torr. We used a combination of $NF_3$ and $CCl_2F_2$ as the reactive gases, each flowing at 5 sccm with an indicated chamber pressure of 30 mtorr; RF power was 150 W, with a plasma voltage of 420 V. The etch rate for $Si$ in these conditions was found to be 1100 nm/min, with quite vertical sidewalls. The etch rate for the $Ni$ mask was negligible for etch times of up to 20 min. Etching was timed so that the etch terminated after slightly etching the buried $SiO_2$ layer, an etch time of 2.5 min. At this point the structures had an appearance as shown in Fig. 1(c).

The $Ni$ mask was removed by wet chemical etching in a mixture of 5 parts nitric acid:5 parts acetic acid:2 parts sulfuric acid:10 parts deionized water (fractions by volume)[6], and rinsing in deionized water. The buried oxide layer was then removed by wet etching in 48 % hydrofluoric acid and rinsing in ethanol. Timing of this step is important, as too short an etch will not free the structures sufficiently, while overetching may remove too much oxide for the structures to remain supported. Etch times of 1.5-3 min. were found to be sufficient. A drawing of the final structure appears in Fig. 1(d).

For larger area suspended structures, we had problems with the suspended $Si$ layer collapsing onto the substrate after removing the samples from the ethanol rinse. This was resolved by transferring the samples, still submerged in ethanol, to a $CO_2$ critical point dryer[7]. The ethanol was flushed away with liquid $CO_2$, at a pressure of about 800 psi, and the $CO_2$ was then heated and pressurized above the critical point (1070 psi, 31.3°C). The dryer was then slowly vented, maintaining the $CO_2$ above the critical temperature. This procedure eliminated almost all of the collapsing observed with air drying of the ethanol. An SEM micrograph of a cradle resonator is shown in Fig. 2(a), where the central beam has dimensions 7.8 μm × 0.5 μm × 0.2 μm. Other structures are shown in Fig. 2(b)-(d).

Using the technique described above, we have fabricated a number of suspended *Si* structures and have measured their resonance properties at a temperature of 4.2 K. Our measurements, which are similar in approach to those of other authors[8], are made by placing the resonant structure in a transverse magnetic field. An applied alternating current is then run through the metal lead on the resonator surface, which drives the resonator by way of the Lorentz force, and the resulting motion of the resonator through the magnetic field generates an electromotive force which is detected by the measurement circuitry. A schematic of the measurement setup is shown in Fig. 3. The radiofrequency drive current is supplied by the source on a network analyzer (HP 3577A), and the signal is lock-in detected by the same analyzer. The calculated resonance frequencies for these structures usually agree with the measured frequencies to within about 10%; in order to find the resonances, a automated search is performed by computer control of the network analyzer.

We have measured the resonance properties of a cradle resonator (see Fig. 2(a)), a torsional oscillator and a tuning fork resonator (Fig. 2(b), (c)), as well as a number of simply suspended beams (Fig. 2(d)). The resonance curve for the cradle resonator is shown in Fig. 4, and for the two tines of the tuning fork in Fig. 5. We have measured a fundamental resonance frequency for a simple beam at 110 MHz, and have fabricated structures with calculated resonance frequencies up to 800 MHz. The typical quality factors that we have measured range from $0.5-2\times10^4$, and are probably limited by losses in the support structure and the metallization layers. Measurements on the higher frequency resonators are in progress, and we are studying methods to increase the quality factors of our structures.

## Acknowledgments

We gratefully acknowledge Dr. Axel Scherer for many valuable conversations, and for use of his reactive ion etch system. This work was supported by DARPA under contract number DABT63-95-C-0112.

> Michael –
> - Revised references
> - Text now references Tortonese's SOI work
> - Figures revised to shorten overall length
>
> A

**Figure Captions**

Fig. 1(a)-(d). Schematic diagram of the processing steps to generate suspended Si structures. The steps are as described in the text.

Fig. 2 (a) SEM micrograph of an undercut Si cradle resonator, where the central beam has dimensions 7.8 μm×0.5 μm×0.2 μm. (b) Torsional oscillator. (c) Tuning fork structure. (d) Simple flexural beams.

Fig. 3. Schematic of the measurement system used to measure mechanical resonances of the Si structures.

Fig. 4. Resonance shape for the central beam of a cradle resonator, with a fit to a Lorentzian lineshape, with resonance frequency and quality factor $f_{res}$ = 15.048 MHz and Q = 4800.

Fig. 5. Resonance shape measured for the two tines of the tuning fork structure.

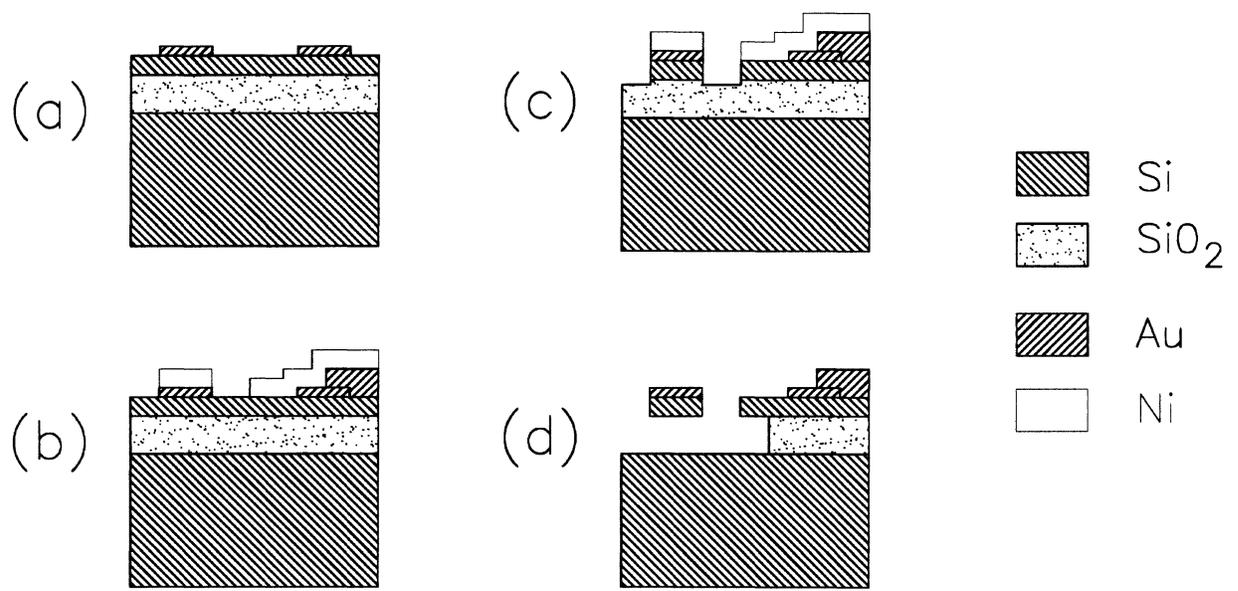

Figure 1.

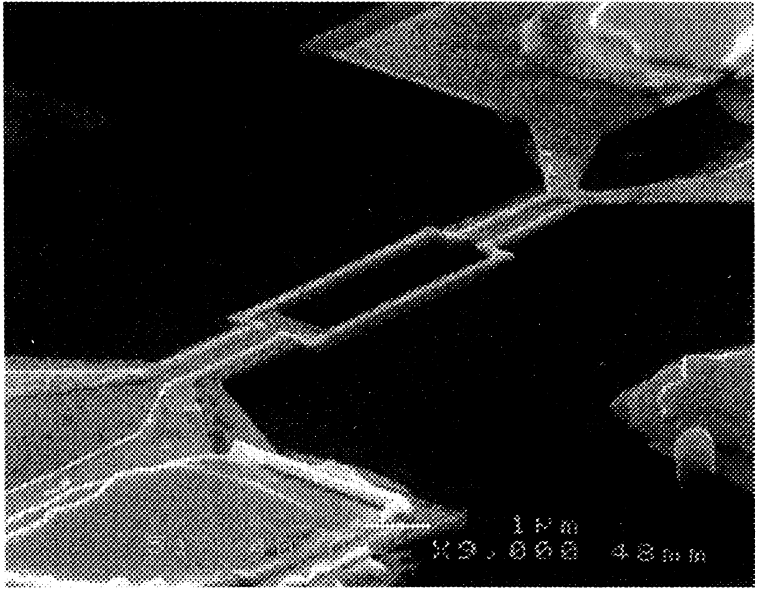 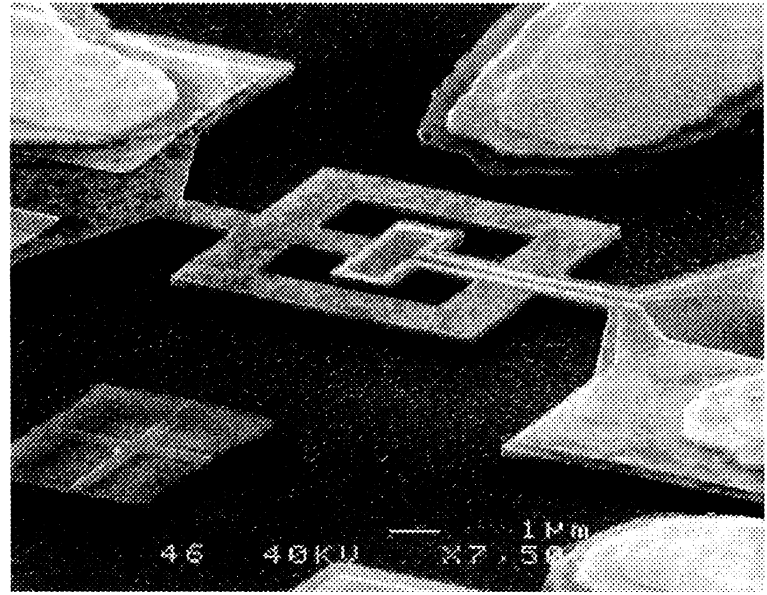
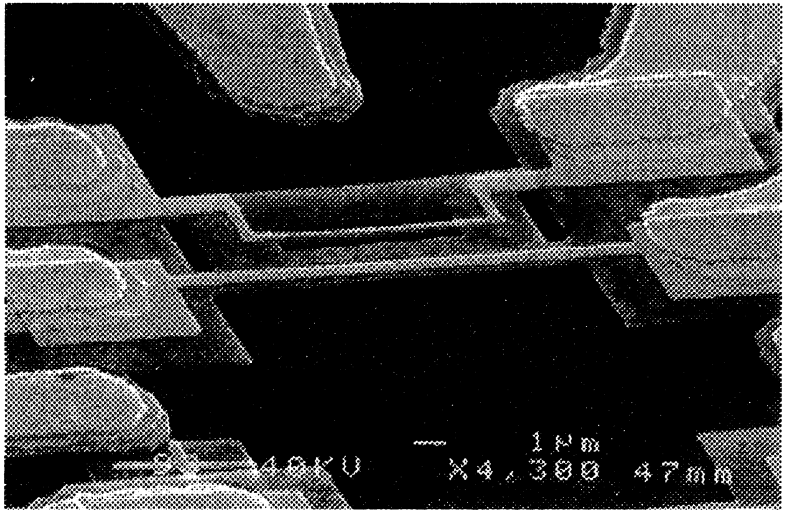 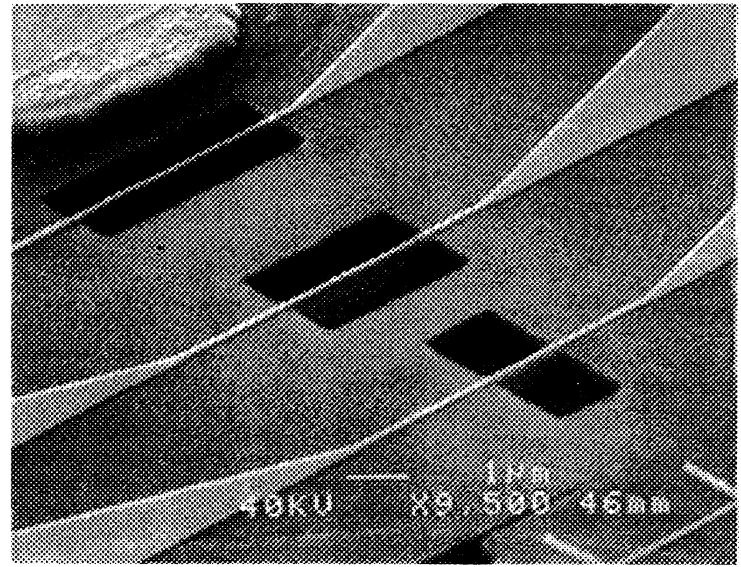

**Figure 2.**

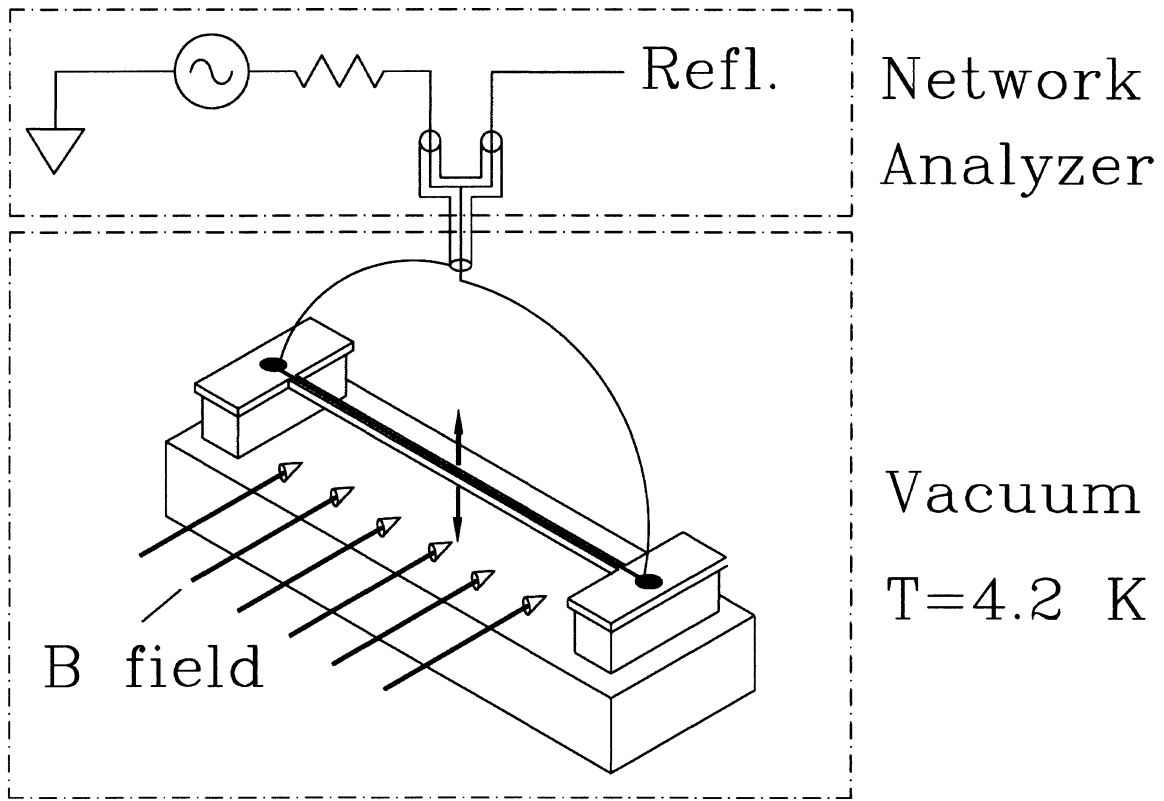

Figure 3.

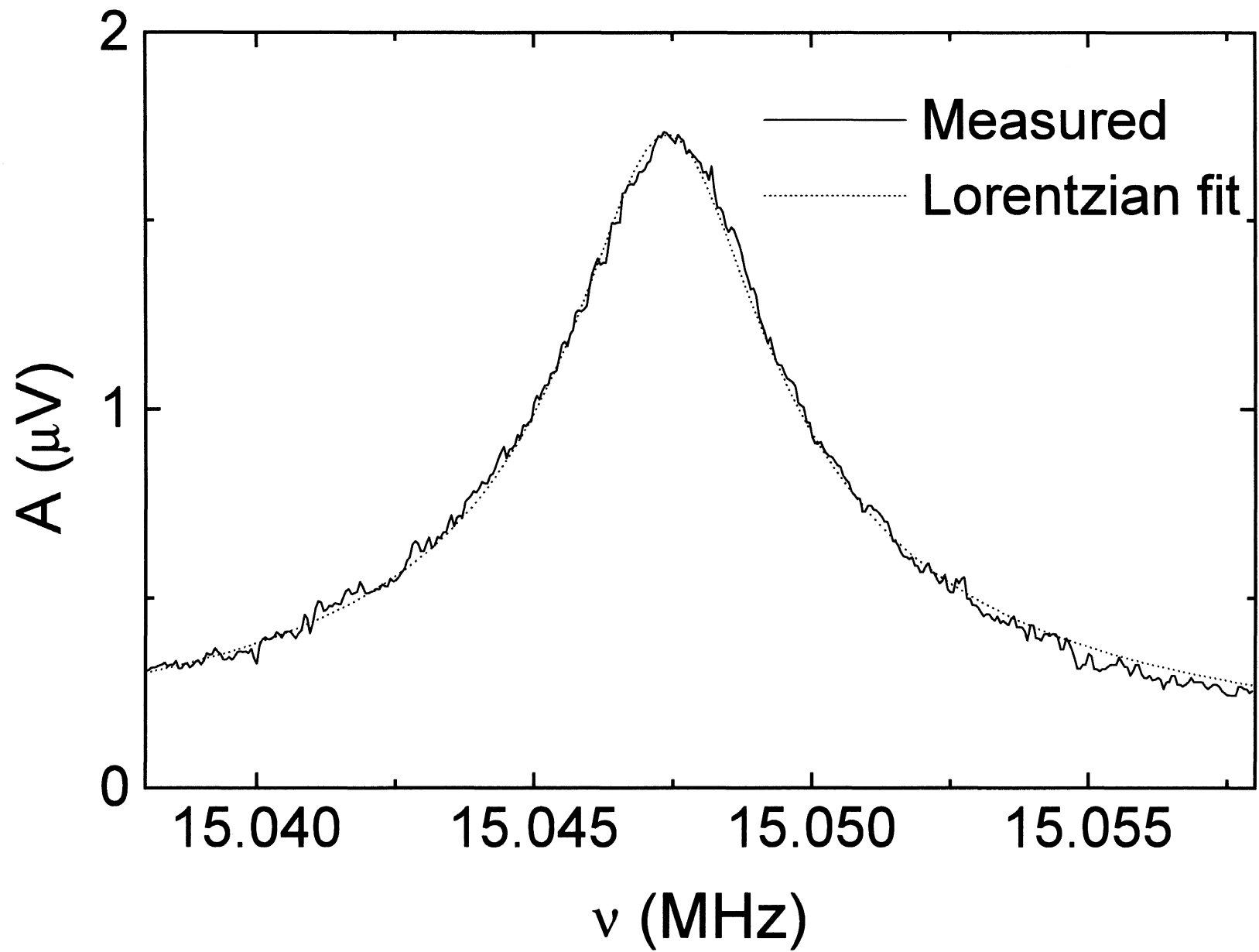

Figure 4.

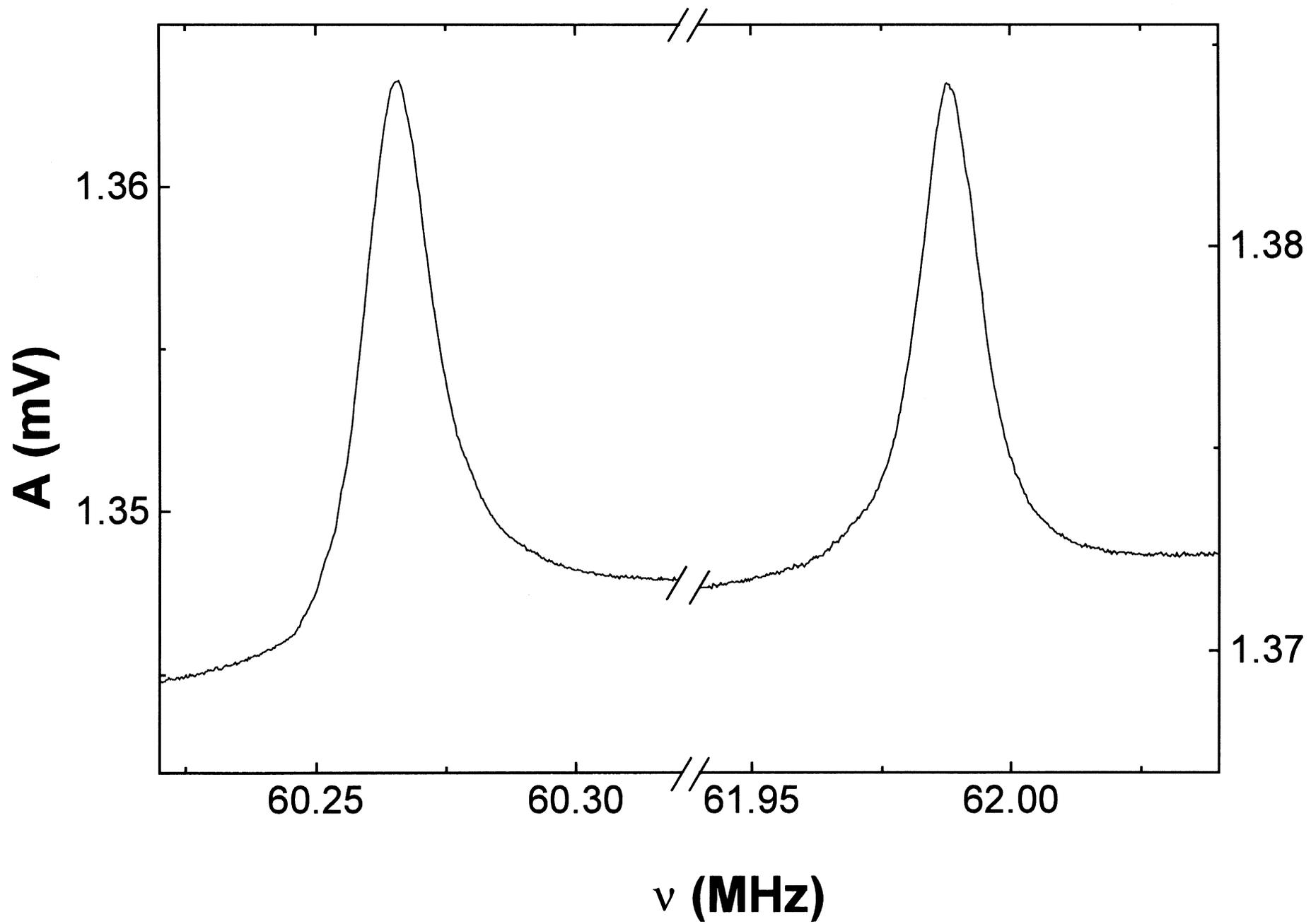

Figure 5.